\documentclass[a4paper]{jpconf}
\usepackage{graphicx}
\usepackage{threeparttable}
\begin{document}
\title{Approaching the event horizon: 1.3mm$\lambda$ VLBI of SgrA*}

\author{Sheperd Doeleman}

\address{MIT Haystack Observatory, Off Route 40, Westford, MA 01886, USA}
\ead{sdoeleman@haystack.mit.edu}

\begin{abstract}
Advances in VLBI instrumentation now allow wideband recording that
significantly increases the sensitivity of short wavelength VLBI observations.
Observations of the super-massive black hole candidate at the center of the
Milky Way, SgrA*, with short wavelength VLBI reduces the scattering effects of
the intervening interstellar medium, allowing observations with angular
resolution comparable to the apparent size of the event horizon of the putative
black hole.  Observations in April 2007 at a wavelength of 1.3mm on a three
station VLBI array have now confirmed structure in SgrA* on scales of just a
few Schwarzschild radii.  When modeled as a circular Gaussian, the fitted
diameter of SgrA* is 37$\mu$as ($+16$,$-10$; $3\sigma$), which is smaller than the
expected apparent size of the event horizon of the Galactic Center black hole.
These observations demonstrate that mm/sub-mm VLBI is poised to open a new
window onto the study of black hole physics via high angular resolution
observations of the Galactic Center.  \end{abstract}

\section{Introduction}
At a distance of $\sim8\;$kpc (\cite{reid93}), SgrA*, the compact radio, NIR
and X-ray source at the Galactic Center, is thought to mark the position of a
super-massive black hole of mass $\sim4\times10^6\;M_\odot$
(\cite{schoedel02},\cite{ghez05}).  Proper motions of SgrA* (\cite{reid04})
confirm that it traces a significant amount of the mass that is inferred by
stellar motions and orbits.  Due to its proximity, SgrA* is the only galactic
nucleus that can be studied with VLBI on sub-AU linear scales
($R_{\mbox{sch}}\sim10\mu\mbox{as}\sim0.1\mbox{AU}$).  The ionized ISM, however,
scatter-broadens images of SgrA* with a $\lambda^2$ dependence, and VLBI at the
highest frequencies is the only available means to set important limits on
intrinsic structures near the event horizon.  VLBI at 7~mm and 3.5~mm has
detected evidence for intrinsic structure of SgrA*, but these observations
remain dominated by scattering effects, and the intrinsic sizes at these
wavelengths (set by the optical depth of the emission) are much larger
than the apparent size of the event horizon (\cite{bower04}, \cite{shen05}).
Evidence from light curves of SgrA* flares from the radio to Xray
(\cite{eckart06}, \cite{yusef-zadeh06}, \cite{marrone08})
implicate structures on smaller ($\sim5-15 R_{\mbox{sch}}$) scales.  Only at
frequencies above 230GHz does the scattering size become smaller than the VLBI
array resolution allowing direct measurement of intrinsic structure on these
scales corresponding to the innermost accretion region.  Over the past decade,
MIT Haystack Observatory has focused on developing next-generation
wideband VLBI instrumentation capable of significantly increasing the
sensitivity of mm/submm VLBI arrays.  The observations described herein used
this instrumentation to show that 1.3mm VLBI of SgrA* can now probe the event
horizon of this super-massive black hole candidate.

\section{Instrumentation}

The focus of new VLBI instrumentation has been to process higher 
bandwidths using commercially available digital technology.  For VLBI, this
has translated into a re-conceptualization of the two main components of 
the traditional VLBI backend: the digitization/formatting stage and the recording
stage.

\subsection{Digital Backend}
\begin{figure}[h]
\begin{center}
\includegraphics[width=20pc,angle=0]{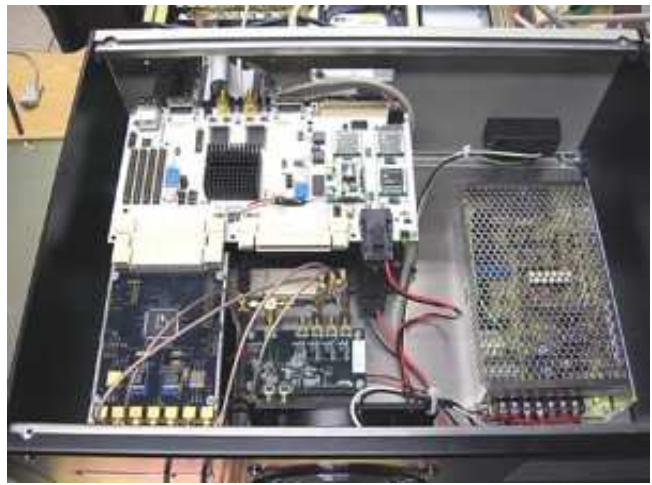}
\caption{\label{dbe_photo} The Digital Backend (DBE) system developed to process two 480MHz wide
IF bands, producing an aggregate VLBI data output rate of 3.84 Gigbit/sec. }
\end{center}
\end{figure}

Prior to recording data for subsequent correlation, the IF (intermediate
frequency) at each VLBI telescope must be sampled and channelized in frequency.
The Mark4 system (similar to the system in use at the VLBA) uses a bank of
analog filters to break the IF into sub-bands.  With the advent of mature FPGA
(Field Programmable Gate Array) technology, it is now feasible to channelize
the IF signal using a polyphase filterbank approach that is realized in digital
signal processing after the sampling stage.  A collaboration between Haystack
Observatory and the Space Science Laboratory at UC Berkeley has focused on
developing a fully digital FPGA-based VLBI backend, which produces an output
suitable for recording on modern hard-disk based recorders.  This system, the
DBE, is capable of sampling two 480MHz wide IFs, and producing two 2Gigabit/sec
output streams (Nyquist sampled, 2-bit resolution).  The DBE (Figure
\ref{dbe_photo}) represents a reduction in backend cost by a factor of $\sim10$
over the previous Mark4 data acquisition system and is 5 times smaller in size
(the size of a single PC).  The next-generation DBE (imaginatively called
DBE2), is currently under development, and will use the Xilinx Virtex5 family
of FPGA chips to double the data throughput to $\sim8$ Gigabits/sec.

\subsection{Mark5 Recorder}
\begin{figure}[h]
\begin{center}
\includegraphics[width=20pc,angle=0]{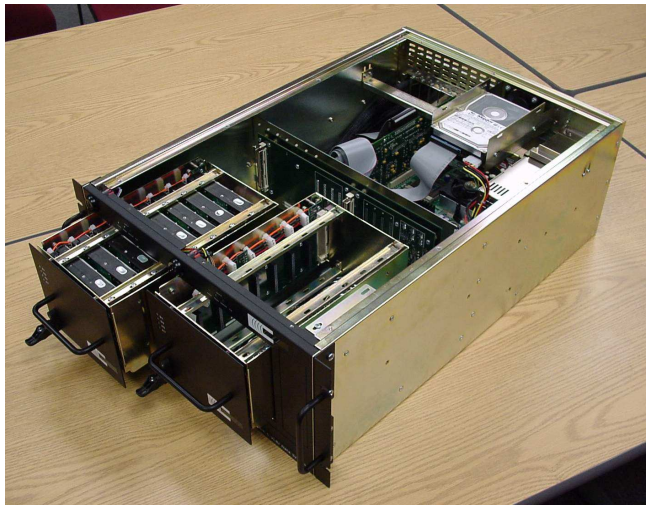}
\caption{\label{mk5_photo} The Mark5 hard disk VLBI data recorder.  Each hard
disk module holds up to 8 individual hard disks.  With current disk sizes, a 
full module can record 6 TBytes.  Maximum recording rates are now 2Gb/s but
will increase to 4Gb/s in 2009.}
\end{center}
\end{figure}

The Mark5 system was developed at Haystack Observatory in collaboration with
Conduant Corp as the first high-data-rate VLBI data system based on
magnetic-disc technology.  Incorporating primarily low-cost PC-based
components, the Mark5 system now supports data rates up to 2048 Mbps, recording
to an array of 8 inexpensive removable hard disks.  The Mark 5 system (Figure
\ref{mk5_photo}) was developed primarily to re-packaged the disks into
convenient '8-pack' modules and over 100 Mark5 units are in use throughout the
VLBI community.  Support for Mark 5 development at MIT Haystack Observatory was
provided by BKG, EVN, KVN, MPI, NASA, NRAO, NSF, and USNO.  The Mark5 system
replaces a magnetic tape system, which used a non-standard reel-to-reel
recorder and a special-purpose tape media whose cost was not likely to decrease
over time.  The new system is over a factor of 5 smaller in size than the
previous recorder with a data rate improvement of x4.  The cost for this new
recording system is $\sim10$ times less than the tape system and uses standard
commercial disk media whose cost (per GigaByte) is projected to substantially
decrease over time.  Efforts to increase recording rates to 4Gigabits/sec are
underway, and the new Mark5C recorder will be capable of recording data from
10Gigabit Ethernet inputs.

\section{Observations}

In April 2007, SgrA* and several quasar calibrators were observed over two
consecutive days at a wavelength of 1.3mm with a three station VLBI array
(\cite{doeleman08a}).  The array included the James Clerk Maxwell Telescope
(JCMT) on Mauna Kea, the Arizona Radio Observatory Submillimeter Telescope
(ARO/SMT) on Mt Graham in Arizona, and one 10m dish of the Coordinated Array
for Research in Millimeter-wave Astronomy (CARMA) in California.  Projected
baseline lengths on SgrA* ranged from $500\times10^6\lambda$ on the shortest
baseline to $3500\times10^6\lambda$ on the longest.  An effective bandwidth of
960 MHz was recorded, resulting in an aggregate recording data rate of 3.84
Gigabits/sec at each site (2 bits/sample, Nyquist sampling).  Data were
processed on the MIT Haystack Observatory Mark4 Correlator to produce complex
visibilities with 0.5 second time resolution.  Calibration quasars were
robustly detected on all three baselines, validating operation of the VLBI
array and allowing refinement of telescope positions for processing of the
SgrA* observations.

\begin{figure}[h]
\begin{center}
\includegraphics[width=40pc,angle=0]{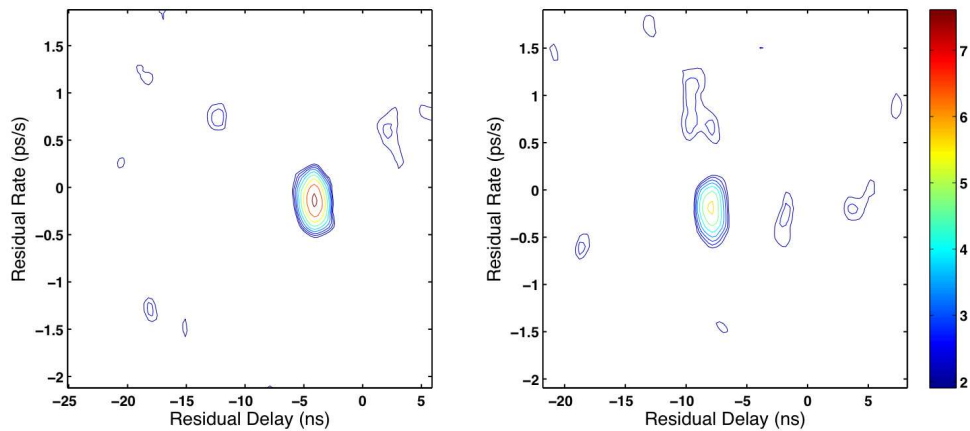}
\caption{\label{detections} Detections of SgrA* and nearby calibrator at 1.3mm$\lambda$
on a 3500km projected baseline between the Submillimeter Telescope on Mt.
Graham, AZ and the James Clerk Maxwell Telescope on Mauna Kea, HI.  Shown are
searches in signal to noise ratio over interferometer delay and delay-rate for
10 minute scans for quasar PKS B1921-293 on April 11, 2007 at 14:00UT (left)
and for SgrA* the same day at 12:00UT (right).  The data were segmented into 8
second intervals to reduce coherence loss due to atmospheric turbulence and the
amplitudes were averaged incoherently.  The formal probability of false
detection (PFD) in each search is computed by comparing the observed signal to
noise ratio with maximal peaks derived from pure noise over the same search
space, and is $<10^{-9}$ for both fringe searches shown above.  Contours in each
plot begin at signal to noise ratio of 2.0 and increase in steps of $2^{1/4}$.  Peak
signal to noise is 7.9 and 5.8 on the left and right searches respectively.}
\end{center}
\end{figure}

Because the geometry of VLBI baselines in an array is not typically known to
$\ll1\lambda$ precision, it is standard practise to search for detections over
a grid of interferometric delay and delay-rate.  A peak in signal-to-noise
ratio of the visibility amplitude, found over a range of Nyquist-sampled delay
and delay-rate space, is deemed a detection if the probability of false
detection is sufficiently low.  At an observing wavelength of 1.3mm,
atmospheric turbulence limits the time over which the VLBI signal can be
coherently integrated.  Therefore, a technique of incoherent averaging
(\cite{rogers95}) was used, to perform the fringe search over each 10 minute
VLBI scan and to determine the VLBI signal amplitude.  Incoherent averaging extends the
effective integration time, but builds signal to noise more slowly than
$\sqrt{t}$.  After measuring the coherence losses due to atmospheric effects
over a range of time scales, the atmospheric coherence time was found to be
$\sim8$ seconds, and the VLBI detection searches were thus made by incoherently
averaging 8 second intervals of coherently averaged data.  These searches
resulted in robust detections and correlated flux density measurements of SgrA*
on both the ARO/SMT-JCMT and ARO/SMT-CARMA baselines (see Figure
\ref{detections}).  No detections were found on the CARMA-JCMT baseline, which
is attributable to the lower sensitivity of that baseline compared with the
others.  The error associated with each visibility amplitude was calculated by
adding in quadrature the noise determined from the detection search with a 10\%
calibration error.  Measurements of SgrA* made with the CARMA array during the
VLBI observations yield a total flux density of SgrA* of $2.4\pm0.25$Jy, which
was observed to be stable over both days, suggesting that SgrA* was observed in
a quiescent state.  Errors in the total flux density measurement are dominated
by pointing and calibration.

\section{Discussion}
\begin{figure}[h]
\begin{center}
\includegraphics[width=20pc,angle=0]{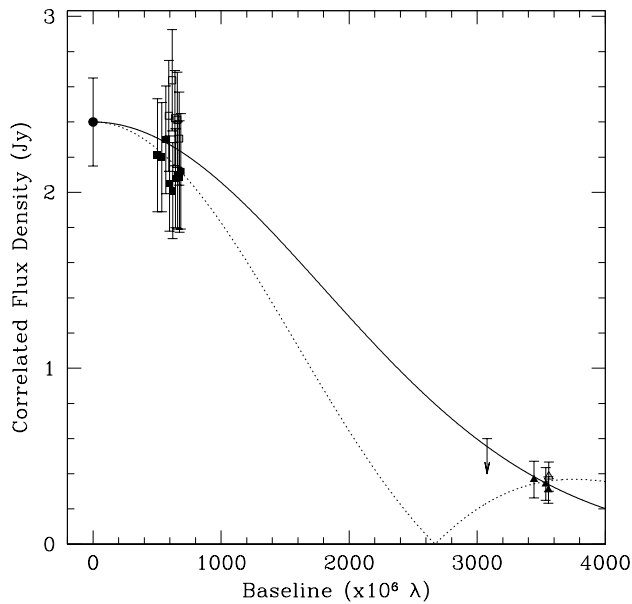}
\caption{\label{uvdist} Shown are the correlated flux density data on the
ARO/SMT-–CARMA and ARO/SMT-–JCMT baselines plotted against projected baseline
length (errors are $1\sigma$). Squares show ARO/SMT-–CARMA baseline data and triangles
show ARO/SMT-–JCMT data, with open symbols for 10 April and filled symbols for
11 April. The solid line shows the weighted least-squares best fit to a
circular Gaussian brightness distribution, with FWHM size of 43.0~$\mu$as. The
dotted line shows a uniform thick-ring model with an inner diameter of 35~$\mu$as
and an outer diameter of 80~$\mu$as convolved with scattering effects due to the
interstellar medium. The total flux density measurement made with the CARMA
array over both days of observing is shown as a filled
circle. An upper limit for flux density of 0.6 Jy, derived from non-detection
on the JCMT-–CARMA baselines, is represented with an arrow near a baseline
length of $3075\times10^6\lambda$.}
\end{center}
\end{figure}

A circular Gaussian model was fit to the VLBI data (shown in Figure
\ref{uvdist}).  The weighted least-squares best-fit model has a total flux
density of $2.4\pm0.5$Jy and full width at half maximum (FWHM) of 43 (+14,-−8)
$\mu$as where errors are $3\sigma$. On the assumption of a Gaussian profile,
the intrinsic size of Sgr A* can be extracted from our measurement assuming
that the scatter broadening due to the ISM adds in quadrature with the
intrinsic size. At a wavelength of 1.3 mm  the scattering size extrapolated
from previous longer-wavelength VLBI (\cite{bower06}) is $\sim22\mu$as.
Removing the scattering effects results in a $3\sigma$ range for the intrinsic
size of Sgr A* equal to 37 ($+16$,$-10$) $\mu$as.  The $3\sigma$ intrinsic size
upper limit at 1.3 mm, combined with a lower limit to the mass of Sgr A* of
$4\times10^5M_\odot$ from measured proper motions yields a lower limit for the
mass density of $9.3\times10^{22}M_\odot\mbox{pc}^{-3}$.  This density lower
limit and central mass would rule out most alternatives to a black hole for Sgr
A* because other concentrations of matter would have collapsed or evaporated on
timescales that are short compared with the age of the Milky Way
(\cite{maoz98}).  

It should be noted, however, that while structure on $4R_{\mbox{sch}}$ scales
is present in SgrA*, models other than the circular Gaussian can be fit to the
data.  This is illustrated by the dotted line in Figure \ref{uvdist}, which
shows the expected flux density as a function of baseline length for a uniform
circular annulus with inner diameter 35$\mu$as and outer diameter 80$\mu$as
that has been scatter broadened by the ISM.  Future higher-sensitivity
observations will distinguish between these two models by allowing detections
of SgrA* on the CARMA-JCMT baseline, which is now represented in Figure
\ref{uvdist} only as an upper limit.

Because of gravitational lensing effects due to the extreme gravity near the
assumed black hole, radiation emitted from near the event horizon of a
non-spinning black hole will have an apparent size of
$3\sqrt{3}R_{\mbox{sch}}$.  For SgrA*, this expected diameter is
$5.2R_{\mbox{sch}}\simeq52\mu$as, which differs by $3\sigma$ from the size
derived from a Gaussian model.  Even if the black hole is maximally spinning
(a=1), the diameter of the event horizon in the equatorial plane ($\sim
45\mu$as) would still exceed the estimated size.  This suggests that SgrA* is
not an optically thick emission region that symmetrically enfolds the black
hole.  Rather, it is likely due either to emission from a jet or from the
approaching (and therefore Doppler enhanced) side of an accretion disk that is
inclined to our line of sight (\cite{falcke00}, \cite{noble07},
\cite{avery06}).  Either scenario results in emission that is offset from the
black hole position.  This marks the first time that astronomical observations
of any kind have directly constrained the spatial relationship between SgrA*
and the black hole.

\section{Conclusions}

The technology to significantly increase the sensitivity of VLBI at wavelengths
of 1.3mm and shorter is now enabling observations of SgrA* on Schwarzschild
radius scales.  Efforts to extend the capabilities of the current 1.3mm VLBI
array include: phasing together radio telescopes on Mauna Kea and at CARMA to
increase collecting area (discussed elsewhere in these proceedings), continuing
to pursue increases in recording bandwidth, and bringing new mm/submm VLBI
sites on-line.  

By 2009, the international collaboration (see the author list of
\cite{doeleman08a}) that has carried out these observations, will field a
higher sensitivity 1.3mm VLBI array that will be sensitive to time variable
structures in SgrA*.  The closure phase is the sum of interferometric phases
around a triangle of baselines, is largely immune from calibration errors, and
deviates from zero in the presence of asymmetric source structure.  Projected
sensitivities will allow monitoring of the closure phase on $\sim10$ second
time scales, and enable tests for periodic structure variation in SgrA* as is
predicted by orbiting hot-spot models of the accretion flow (\cite{avery06},
\cite{genzel}).  By timing periodicities in the closure phase, one can extract
the fundamental black hole spin parameter (\cite{doeleman08b}).  Thus, by
spatially resolving the innermost accretion region surrounding SgrA*, mm/submm
VLBI is now positioned to address fundamental issues in black hole physics.

\section*{Acknowledgments}
VLBI at mm/submm wavelengths would not be possible without the dedicated support of
staff and scientists as all participating facilities.  VLBI work at the MIT Haystack
Observatory is supported through grants from the National Science Foundation.

\end{document}